# Tuning the electronic band structure in a kagome ferromagnetic metal via magnetization

Neeraj Kumar[1] and Y. Soh[1*],

1. Paul Scherrer Institut, 5232 Villigen, Switzerland

Yihao Wang[2], Junbo Li[2], and Y. Xiong[2]
2. Anhui Province Key Laboratory of Condensed Matter Physics at Extreme Conditions, High Magnetic Field Laboratory of the Chinese Academy of Sciences, Hefei 230031, China

**Abstract:**
Materials with zero energy band gap display intriguing properties including high sensitivity of the electronic band structure to external stimulus such as pressure or magnetic field. An interesting candidate for zero energy band gap are Weyl nodes at the Fermi level $E_F$. A prerequisite for the existence of Weyl nodes is to either have inversion or time reversal symmetry broken. Weyl nodes in systems with broken time reversal symmetry are ideal to realize the tunability of the electronic band structure by magnetic field. Theoretically, it has been shown that in ferromagnetic Weyl materials, the band structure is dependent upon the magnetization direction and thus the electronic bands can be tuned by controlling the magnetization direction. Here, we demonstrate tuning of the band structure in a kagome Weyl ferromagnetic metal $Fe_3Sn_2$ with magnetization and magnetic field. Owing to spin-orbit coupling, we observe changes in the band structure depending on the magnetization direction that amount to a decrease in the carrier density by a factor of four when the magnetization lies in the kagome plane as compared to when the magnetization is along the *c* axis. Our discovery opens a way for tuning the carrier density in ferromagnetic materials.

## Introduction:

The concept of tuning the electronic properties, such as the carrier density, in semiconductors by an electric field was conceived[1] long before a practical field-effect transistor (FET)[2] was realized, which brought unforeseen applications and technical advancements. Much harder is to tune the electronic properties of a metal with an electric field due to the large carrier density and accompanying efficient screening of the electric field. Therefore, an alternative method is necessary in order to tune the carrier density in metals. An option is to use a magnetic field instead of an electric field to tune the electronic properties of a metal since a magnetic field is not screened in a normal metal, and take advantage of the spin-orbit coupling, which dictates that the electronic structure depends on the spin or magnetization. While the concept is attractive, the demonstration to achieve large carrier density modulation with a magnetic field is still lacking. In most metals, the effect of the change of the band structure via magnetization is not sufficient to yield an appreciable effect. However, it can be realized when the band structure is such that the Fermi surface is near a saddle point of the band dispersion such as in ferromagnet $ZrZn_2$[3] or in ferromagnetic Kondo lattice $YbNi_4P_2$[4], in which cases the Fermi surface topology changes between a connected and disconnected Fermi surface around the saddle point by changing the magnetic field strength resulting in a Lifshitz transition[5]. In $WTe_2$ modulation in the pockets sizes is also seen with the magnetic field strength[6]. Such Lifshitz transitions were accompanied by metamagnetic transitions. Instead of changing the magnetic field strength, Liftshitz transitions can be enabled in ferromagnetic Weyl semimetals by rotating the magnetization [7, 8] since the band structure and location of Weyl nodes can be controlled by the magnetization direction. Manipulation of the Fermi surface in a prototypical ferromagnet iron has been demonstrated by angle resolved photoemission spectroscopy (ARPES)[9], but measurement of the tuning of the carrier density by magnetic field or magnetization in metals is lacking.

$Fe_3Sn_2$ has been shown theoretically to be a Weyl semimetal with Weyl points moving depending on the magnetization direction[10] due to the band structure being dependent upon the magnetization direction. Recently, it was also shown by band structure calculations in the case of hcp-Co and Weyl semimetal $Co_3Sn_2S_2$ that it is possible to tune the Weyl point to the Fermi level $E_F$ using magnetization[11]. Tuning of the band topology with the magnetization vector was also recently shown in $Co_2MnAl$, where different values of the anomalous Hall effect was measured for magnetization along different directions[12]. While very large fields (23 Tesla at 2 K) are required in the case of $Co_3Sn_2S_2$ to rotate the magnetization from the easy axis *c* to the *ab* plane[13], in the case of $Fe_3Sn_2$, the magnetic anisotropic energy is very low[14] and a field of about 1 T is sufficient to fully rotate the magnetization along any direction[15, 16]. Thus, $Fe_3Sn_2$ is an ideal case to investigate the tuning of the band structure using magnetization direction as a control knob[17].

We investigate the Hall effect as a function of temperature, field, and magnetization angle, and discover a change in the carrier density as a function of magnetization angle at low temperature. The change in carrier density with magnetization angle is possible only in the low temperature phase where the easy axis lies in the *ab*-plane and not in the high temperature phase where the easy axis is along the *c* axis. Thus, in addition to the change in the easy axis[18], the spin reorientation transition is also accompanied by a fundamental change in the electronic properties of $Fe_3Sn_2$[15, 16]. This band tuning starts only below 80 K, which we have earlier reported to be the temperature at which the spin reorientation to the low temperature phase is complete with the magnetization of the entire sample being aligned in the *ab* plane[16].

**Experimental Methods:**

$Fe_3Sn_2$ consists of kagome lattices of Fe with Sn at the center as shown in the right of Fig. 1(a), stacked along the *c* axis as bilayers and separated by Sn layers as shown in the left of Fig. 1(a). The crystal structure belongs to the R-3m space group. A single crystal of $Fe_3Sn_2$ grown by vapour transport method was used to investigate the angular dependent Hall effect. Due to the crystal geometry being flat and extended in the kagome plane and thin along the *c* axis, we chose a measurement geometry such that the current is applied along the *a* axis in the kagome plane with the Hall voltage probes perpendicular to the current and in the kagome plane. The magnetic field $\vec{H}$ is applied in different directions to study the effect of the magnetization direction on the electronic band structure of $Fe_3Sn_2$.

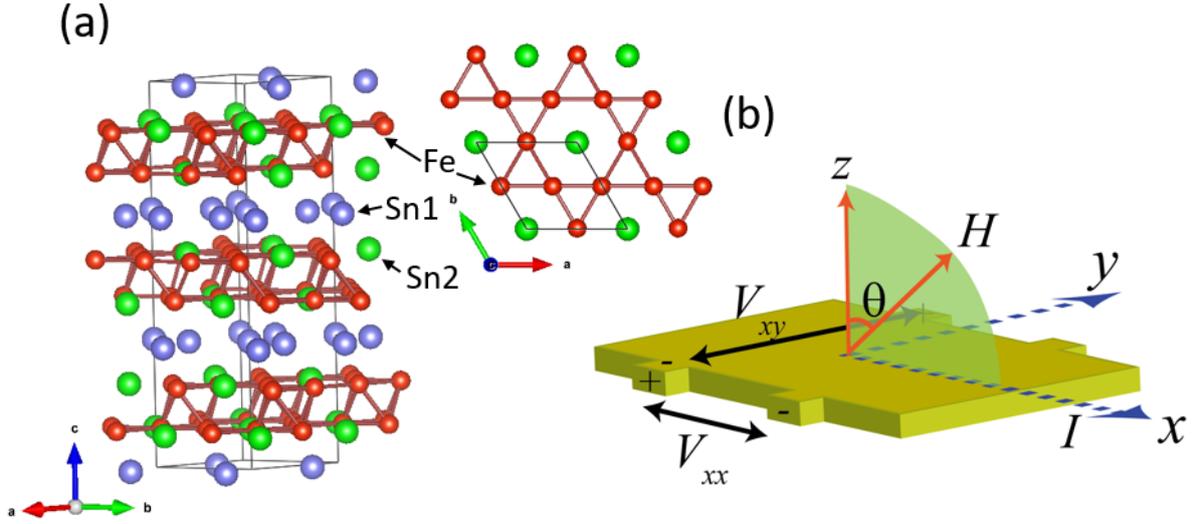

Figure 1 (a) Crystal structure of $Fe_3Sn_2$. (b) Schematic diagram of the measurement setup and direction of field rotation.

**Results and Discussion:**

Section 1: Behavior at high temperatures

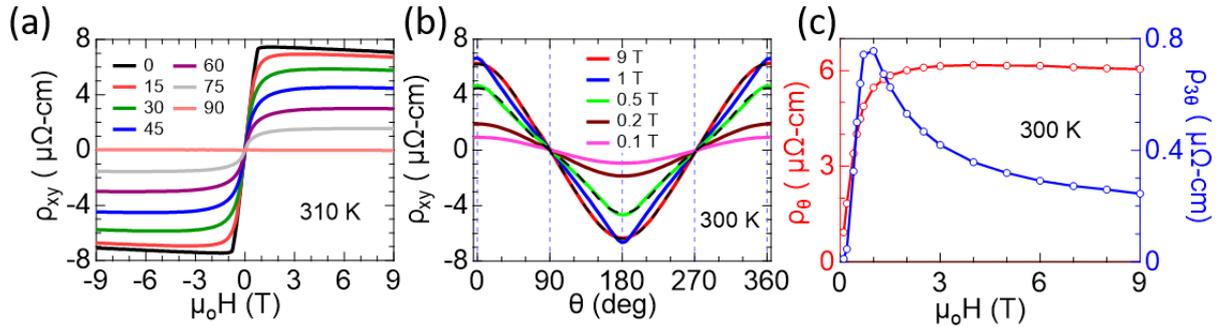

Figure 2 (a) Hall effect at 310 K for field along several directions. 0 deg is H//001. (b) Angular dependence of the Hall effect at selected magnetic field values at 300 K. (c) Dependence of $\cos\theta$ and $\cos3\theta$ terms on the magnetic field at 300 K.

First, we look at the Hall effect at room temperature. Fig. 2(a) shows the Hall effect of $Fe_3Sn_2$ at 300 K for field along several directions. 0° refers to $\vec{H} \parallel \mathbf{c}$ and 90° to $\vec{H} \parallel \mathbf{a} \parallel \vec{I}$, respectively as shown in Fig. 1(b). The curves display a behavior typical of ferromagnetic materials where an ordinary Hall effect (OHE) due to the carrier density and an anomalous Hall effect (AHE) due to the magnetization is observed. The total Hall resistivity is $\rho_{xy}(H) = R_H H_c + R_A M_c$, where $H_c$ is the z component of the external magnetic field and $M_c$ is the z component of the magnetization. The carrier density obtained from the fitting of the 0° curve above 1 T is about $1.1 \times 10^{28}$ carriers/m³, which together with the measured AHE magnitude, is in good agreement with previous reports[19, 20]. Upon rotating the magnetic field, $H_c$ and $M_c$ decreases and thus the Hall magnitude is reduced. Fig. 2(b) shows the variation of the Hall resistivity $\rho_{xy}$ as the magnetic field is rotated in the *ac* plane for several values of the magnetic field. $\rho_{xy}$ decreases with angle to zero when the current and magnetic field become parallel. The OHE depends on the out of plane component of the magnetic field perpendicular to the plane, i.e., $H_c$, and thus should vary as $cos\theta$, while the AHE depends on the perpendicular component of the magnetization $M_c$. At low field, the magnetization vector is not parallel to the magnetic field due to the magnetocrystalline anisotropy. At high magnetic field, such as 9 T, $\vec{M} \parallel \vec{H}$, thus $\rho_{AHE}$ should also follow $cos\theta$. Instead we find that $\rho_{xy}$ follows the equation

$$\rho_{xy} = \rho_\theta cos\theta + \rho_{3\theta} cos3\theta.$$

The fitting for 0.5 T and 9 T data in Fig. 2(b) with the above equation shows an excellent fitting. Close to the saturation field, however, the fittings slightly deviate due to the magnetocrystalline anisotropy. The fitting can be improved with the introduction of higher order terms up to $cos7\theta$, but we limit up to $cos3\theta$ for the analysis. Fig. 1(c) shows the dependence of the parameters $\rho_\theta$ and $\rho_{3\theta}$ on the magnetic field at 300 K. The $\rho_\theta$ term increases sharply with field up to 1 T, due to the main contribution from the AHE, which saturates at $M_s$, and continues increasing slowly up to 3 T before decreasing slightly at higher fields. The $\rho_{3\theta}$ term also increases sharply with field up to 1 T but decreases much more rapidly with the field after 1 T. The angular dependence of the Hall effect in $La_{0.8}Sr_{0.2}MnO_3$ films was also found to follow a similar behavior[21]. It was suggested that the $\rho_{3\theta}$ term either originates from an in-plane magnetic anisotropy or from the dependence of the AHE coefficient $R_A$ on the longitudinal resistivity $\rho_{xx}$, which itself changes as $cos2\theta$. In our system, $\rho_{xx}$ also displays an angular dependence of $cos2\theta$, which together with $cos\theta$ term give a $cos3\theta$ dependence. Perhaps some unsaturated magnetic domains in $Fe_3Sn_2$ remain even at 9 T which are responsible for the $\rho_{3\theta}$ term. Interestingly both the $\rho_{3\theta}$ term and anisotropic magnetoresistance (AMR) are observed to decrease with increasing field. Moreover, we found the dependence of $\rho_{xy}$ on the direction of $\vec{H}$ to be nearly isotropic in the basal plane. The angular dependence was found to be the same for rotation of the magnetic field either in the $ac$ plane or in the $a_\perp c$ plane (where $a_\perp$ is a vector perpendicular to $a$ in the $ab$ plane), except for a small difference arising from the planar Hall effect. More details are provided in the Supplemental Material (section 2).

Section 2: Behavior at low temperatures

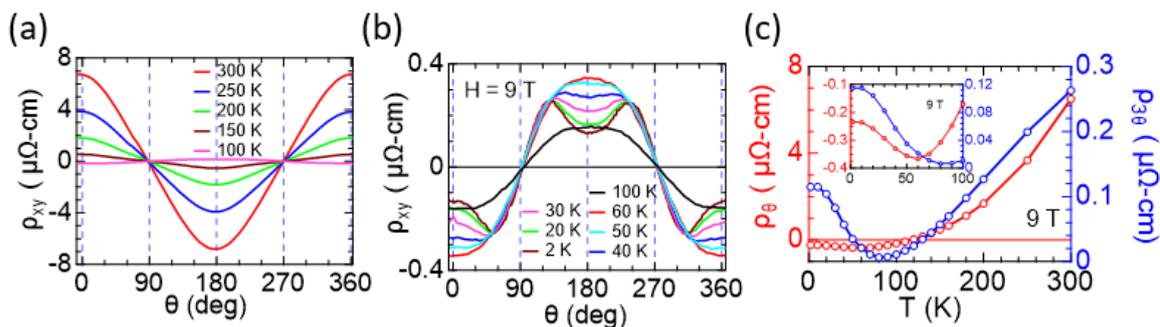

Figure 3 Angular dependence of the Hall effect in the temperature range (a) 300 to 100 K and (b) 100 to 2 K. (c) variation of coefficients for $cos\theta$ and $cos3\theta$ term with temperature at 9 T. Inset: low temperature variation.

Fig. 3(a) shows the angular dependence of the Hall effect at 9 T in the temperature range 300 to 100 K. It shows qualitatively a similar behaviour from 300 down to 150 K. The magnitude of the Hall effect decreases as expected and changes sign becoming negative at 100 K. This is due to a competition between the positive AHE and the negative OHE. The AHE decreases rapidly with temperature and becomes negligible below 100 K, while the changes in the OHE are non-monotonic. Fig. 3(b) shows the angular dependence of the Hall effect at 9 T in the temperature range of 100 K and below. Clearly the sign of the Hall effect at $\theta = 0$, where $\vec{H} \parallel c$, is reversed compared to the high temperature phase of 150 K and above. In addition, below 50 K, a reversal of trend in $\rho_{xy}$ is seen with a decrease in the magnitude of the Hall signal at 0° and 180° in contrast to the 60-100 K range where the magnitude of the Hall signal increases as the temperature decreases. Moreover, below 50 K, the angular dependence of $\rho_{xy}$ is qualitatively different from the high temperature phase. Instead of an angular dependence resembling a sinusoidal function, $\rho_{xy}$ shows extrema at angles away from the normal direction to the Hall bar. These changes are reflected in the $\rho_\theta/\rho_{3\theta}$ curves as shown in Fig. 3(c). Both $\rho_\theta$ and $\rho_{3\theta}$ decrease with decreasing temperature until reaching the minima at 60 K and 80 K, respectively. While at room temperature $\rho_\theta$ is much larger than $\rho_{3\theta}$, at low temperature these two quantities are comparable. $\rho_\theta$ crosses zero around 120 K, which accounts for the change of sign of $\rho_{xy}$ as the system is cooled down. This coincides with the spin reorientation transition temperature. The strong deviation from a sinusoidal angular dependence at low temperature is reflected in the increase of $\rho_{3\theta}$ below 80 K. It is worth noting that $\rho_{3\theta}$, which accounts for a deviation from the simple picture that the angular dependence of $\rho_{xy}$ only depends on the perpendicular components of $\vec{M}$ and $\vec{H}$, is more dominant at low T when considering its magnitude relative to $\rho_\theta$.

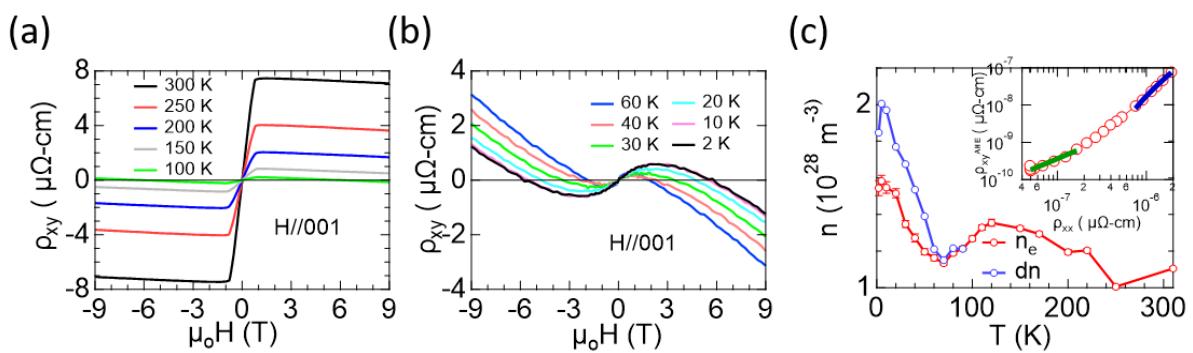

Figure 4 Hall effect in the temperature range (a) 300 to 100 K and (b) 60 to 2 K, (c) variation of the carrier density and AHE coefficient with temperature. Inset: the AHE resistivity as a function of longitudinal resistivity.

In order to understand the origin of the non-sinusoidal angular dependence of $\rho_{xy}$ at low temperatures, we studied the field dependence of $\rho_{xy}$ as a function of temperature and field direction. We first focused on the Hall effect in the perpendicular configuration such that $\vec{H} \parallel c$. Fig. 4(a) and (b) shows the Hall effect in the temperature range 300 to 2 K. The overall Hall effect magnitude decreases and becomes negative below 100 K at a field of 9 T. After a maximum negative value at 60 K, the Hall effect at 9 T stays negative but decreases in magnitude at lower temperatures. This is due to a competition between the positive AHE and the negative OHE, both of which are temperature dependent. The AHE deceases rapidly with temperature and becomes negligible below 100 K. Analysis of the AHE with $\rho_{xy} \propto \rho_{xx}^{power}$ (inset of Fig. 4(c)), reveals a quadratic dependence above 150 K and a linear dependence below 60 K suggesting a change in the main mechanism from extrinsic side jump or intrinsic Kapplus-Luttinger (KL) mechanism at high temperature to skew scattering at low temperature. This is another signature of the spin reorientation transition and accompanying change of the electronic structure. It is to be noted that in Wang et al. the change in the exponent was much smaller[19].

Now, we examine the change in the carrier concentration. As shown in Fig. 4(a), at high temperatures, the Hall effect is linear above the saturation field, and thus the carrier density is obtained by using the linear slope at these temperatures. A negative slope suggests electrons to be the majority carrier, which agrees with previous reports. While the data at high temperature gives a good fit to $\rho_{xy}(H) = R_H H_c + R_A M_c$, at low temperature, the data (Fig. 4(b)) does not fit to this equation. Instead, we consider adopting a two-carrier model consisting of an electron term and a hole term, together with an anomalous Hall effect term. In such a case, the Hall effect is given by

$$\rho_{xy}(H) = \frac{H}{e} \frac{(n_h \mu_h^2 - n_e \mu_e^2) + (n_h - n_e)\mu_e^2 \mu_h^2 H^2}{(n_e \mu_e + n_h \mu_h)^2 + [(n_h - n_e)\mu_e \mu_h]^2 H^2} + R_A M_Z$$

where $n_e, n_h$ are the concentration and $\mu_e, \mu_h$ are the mobility of electron and holes, respectively. The qualitative conclusions are, however, unaffected by the model used. The carrier density obtained in a single carrier model, where a linear fit for $|\vec{H}| \geqq 6$ T is used, is equivalent to the net difference of the carrier densities obtained in the two-carrier model, where the data for $|\vec{H}| \geqq 1$ T was fit. First, let us discuss the results obtained from the single carrier picture. The electron carrier density monotonically increases from about $1.1 \times 10^{28}\ m^{-3}$ at 300 K to about $1.4 \times 10^{28}\ m^{-3}$ at 120 K, below which first a small decrease and then a big jump is observed in the carrier density as shown in Fig. 4(c). At 2 K, the carrier density is $1.8 \times 10^{28}\ m^{-3}$. While the trend of the carrier density above 120 K agrees with the literature, the behavior below 120 K is different from the results by Wang et al. where the carrier density decreases at low temperature[19]. We found the behavior below 120 K to be sample dependent with the carrier concentration either decreasing or increasing (see Supplemental Material).

In the two-carrier model, a simultaneous fitting of the magnetoresistance (MR) and the Hall effect is required for an accurate assessment of the electron and hole carrier densities and mobilities. Unfortunately, only the Hall effect data gives a good fitting and the magnetoresistance data does not give a good fitting with any set of parameters (see Supplemental Material). Fitting with only the Hall effect can, however, accurately determine the difference $\Delta n$ in the carrier densities. As shown in Fig. 4(c), $\Delta n$ follows the same behavior as n, which slightly overestimates the net carrier density. The difference in n and $\Delta n$ increases as the temperature is lowered as the field dependence of $\rho_{xy}$ deviates further from linear.

Within the two-carrier model, at 2 K, looking at the negative slope of $\rho_{xy}(H)$ at the high field limit, we conclude that $n_e > n_h$, while from the positive slope of $\rho_{xy}(H)$ at the low field limit, $\mu_e < \mu_h$. A large range of values for $n_e, n_h, \mu_e, \mu_h$ fits the Hall effect data, as the Hall effect alone cannot determine all four quantities. In order to make an estimate, we use an extra constraint. The first option we try is to fit the MR with a second order polynomial and use the coefficient of $H^2$ as $\mu_e \mu_h$. This yields a very small $n_h = 5.5 \times 10^{25} m^{-3}$ but large $n_e = 1.75 \times 10^{28} m^{-3}$, while at the same time a very large $\mu_h = 1.8 \times 10^{-1} m^2 V^{-1} s^{-1}$ but small $\mu_e = 7.5 \times 10^{-3} m^2 V^{-1} s^{-1}$. $\rho_{xx}(0)$ in this case is $4.3\ \mu\Omega.cm$ which is close to the experimental value of $\rho_{xx}(0) = 4.35\ \mu\Omega.cm$. Fitting the MR with a power law $H^p$ and using the coefficient of $H^p$ as $(\mu_e \mu_h)^p$, yields $\rho_{xx}(0) = 2.27\ \mu\Omega.cm$, $n_h = 1.69 \times 10^{26} m^{-3}$, and $\mu_e = 1.4 \times 10^{-2} m^2 V^{-1} s^{-1}$, while $n_e, \mu_h$ remain almost the same as before. Thus, it is difficult to obtain correct values for the carrier densities and mobilities with the current data.

In the above cases, the hole pocket is very small, roughly 2-3 orders of magnitude smaller than the electron pocket. On the other hand, the hole mobility is 1-2 orders of magnitude larger than the electron mobility. We find these fitting parameters unphysical.

If we force the electron and hole pockets to be of comparable magnitude but consistent with the measured $\Delta n$ value, the resulting longitudinal resistivity that we obtain is very small. In this scenario, $n_h = 1 \times 10^{28} m^{-3}$, $n_e = 2.74 \times 10^{28} m^{-3}$, $\mu_e = 1.4 \times 10^{-1} m^2 V^{-1} s^{-1}$, and $\mu_h = 7.3 \times 10^{-1} m^2 V^{-1} s^{-1}$ is obtained, which results in a very small value for $\rho_{xx}(0) = 0.055\ \mu\Omega.cm$, two orders of magnitude smaller than the experimental value. This again is unphysical showing that at $T \lesssim 100$ K, the Hall effect even with $\vec{H} \parallel c$ is unconventional. More combination of fitting parameters are shown in Supplemental Material Fig. S4 and S5 along with the fitting curves for the above parameters.

In order to understand the angular dependence of the Hall effect at low temperature deviating from a sinusoidal behavior, we studied the field dependence of $\rho_{xy}$ with $\vec{H}$ oriented in different directions. As shown in Fig. 3(b) and 5(b), the Hall effect magnitude at 9 T for $T \lesssim 40$ K initially *increases* upon rotating the field away from the out of plane axis before decreasing and eventually crossing zero at $\theta = 90°$. As previously shown, the AHE at 2 K is almost negligible as compared to the OHE. The OHE magnitude depends only on the out of plane component of the magnetic field $H_z$, which decreases as the field vector is rotated away from the *c* axis to the *a* axis. In order to elucidate how peculiar is the measured angular dependence of $\rho_{xy}(\theta)$, in Fig. 5(b), we plotted what $\rho_{xy}$ would be (blue line) if it only depended on the out of plane component of the magnetic field $H_z$ using the field dependent $\rho_{xy}(H)$ data obtained with $H \parallel c$. The measured $\rho_{xy}$ value clearly deviates from what $\rho_{xy}$ would be if only the out of plane component of the magnetic field was relevant.

To understand this unusual behavior, where the magnitude of the OHE initially increases as $H_z$ decreases, we looked at the field dependence of the Hall effect at several angles (see Fig. 5(a)) at 2 K, where the effect is the strongest. We plotted the dependence of the Hall effect on the out of plane component of the magnetic field $H_z$. If the Hall effect only depended on $H_z$, all the curves would collapse. However, they peel off from each other suggesting that the band structure is different depending on the field direction. We find that such an unusual Hall effect behavior could be explained by assuming a carrier density modulation by the magnetization direction. Band structure modulation with the magnetization has already been shown theoretically and by ARPES on pure Fe[9]. Fe$_3$Sn$_2$ is a soft ferromagnet and thus the magnetization and the field are parallel to each other at fields such as 9 T, which is very high as compared to the saturation field for any direction[16]. Using the two-carrier model, we found that $\Delta n$ decreases as the field is rotated towards the kagome plane (Fig. 5(c)). A one-carrier model with a linear fitting at high fields $|\vec{H}| \geqq 6$ T also results in the same conclusion (see Supplemental Material). Reduction in $\Delta n$ with angle points towards increasing compensation as the magnetization lies in the kagome plane. A reduction of $\Delta n$ by a factor of four is observed at 2 K in the current experiment. This result is verified by measuring two more samples where we see qualitatively the same behavior (see Supplemental Material). Ye at al. have reported changes in the dHvA oscillation frequency in Fe$_3$Sn$_2$ upon rotation of the magnetic field more rapidly than they expected, which required to consider the spin-orbit coupling effect that would yield an evolving band structure depending on the magnetization direction [22]. We believe that both of these results have a common origin.

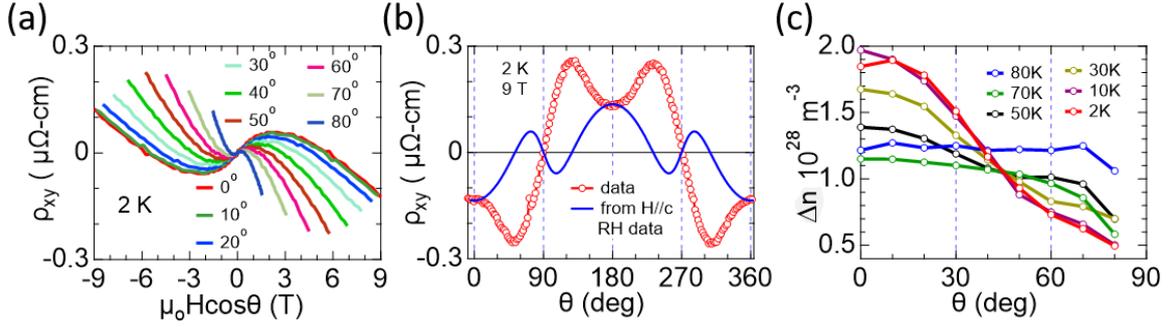

Figure 5 (a) Hall effect at 2 K at several angles as a function of out of plane field component. (b) Angular dependence of the Hall effect at 2 K 9T (red circles). (c) Δn vs angle at several temperatures.

At room temperature no unusual angular dependence of $\rho_{xy}$ or associated changes in *n* is observed as the magnetization vector is rotated as shown in Fig. 2(b). Thus, we probed this unusual behavior as a function of temperature. We found that the change in the carrier density with magnetization rotation, starts to occur only below 80 K. At 80 K, we do not see any clear evidence of carrier density modulation with magnetization rotation while at 70 K a change in carrier density from $1.15 \times 10^{28} m^{-3}$ to $0.58 \times 10^{28} m^{-3}$, which is about a factor of two (Fig. 5(c)), is observed. Upon cooling, the change in the carrier density increases up to a factor of four at 2 K. More data demonstrating the change in the Hall resistivity with angle is shown in the Supplemental Material. Fig S2 shows data at 2 K and various magnetic field values. Fig. S6 shows the change in the carrier density with magnetization vector in a one band picture as well.

In the same temperature range of $T < 80$ K where the band structure is sensitive to the magnetization direction, we noticed the failure of the two-carrier model in explaining the nonmonotonic behaviour of $\rho_{xy}(H)$ with respect to H for $\vec{H} \parallel c$. Given that the electronic band structure is sensitive to the direction of $\vec{M}$ at $T < 80$ K, it is quite likely that it is also sensitive to $\vec{H}$. It is known that in zero-gap materials where the conduction band and valence band edge meet at the Fermi level, there is no threshold to move electrons from occupied states in the valence band to empty states in the conduction band, and therefore the band structure is very sensitive to external factors such as pressure or magnetic field. We believe that we have zero-gap bands, in addition to conduction bands, by having Weyl nodes near $E_F$. The nonmonotonic behaviour of $\rho_{xy}(H)$ with respect to H could be due to the presence of Weyl nodes near $E_F$. The sensitivity of the anomalous Hall conductivity to the position of $E_F$ with respect to the Weyl nodes have been underexplored. Recent theoretical calculations have shown that the anomalous Hall conductivity displays a dome shape as the position of $E_F$ is varied with respect to the Weyl nodes[23]. Our observation of a dome shaped $\rho_{xy}(H)$ as we vary H could be due to the energy shift of the Weyl nodes with respect to $E_F$ due to the Zeeman effect.

**Conclusion:**
We have demonstrated modulation of the electron band structure and carrier density below 80 K in a kagome Weyl ferromagnet $Fe_3Sn_2$ via rotation of the magnetization direction. At the same temperature range where the band structure is sensitive to the magnetization direction, a nonmonotonic behaviour of $\rho_{xy}(H)$, which cannot be explained by a two-carrier model, is observed. Both effects can be explained by the presence of zero-band gap, *i.e.* Weyl nodes, at the Fermi level, which are sensitive to the magnetization due to spin-orbit coupling and the strength of the magnetic field due to Zeeman effect.

**Methods:**

Details of the sample preparation and measurements are the same as in our previous paper on anisotropic magnetoresistance in $Fe_3Sn_2$[16].

# Tuning the electronic band structure in a kagome ferromagnetic metal via magnetization

Neeraj Kumar[1] and Y. Soh[1*],

1. Paul Scherrer Institut, 5232 Villigen, Switzerland

Yihao Wang[2], Junbo Li[2], and Y. Xiong[2]

2. Anhui Province Key Laboratory of Condensed Matter Physics at Extreme Conditions, High Magnetic Field Laboratory of the Chinese Academy of Sciences, Hefei 230031, China

**Supplemental Material**

**Section 1: Field dependence of $\cos 3\theta$ term associated with anisotropic magnetoresistance**

The $\cos 3\theta$ term decreases with the magnetic field at room temperature as shown in Fig. 2(c) in the main text. We observe that the anisotropic magnetoresistance (AMR) ($\rho_c - \rho_a$) also decreases with the field above the saturation field[1]. Fig. S1 shows the dependence of these two terms with the magnetic field at 300 K. It could be that these two quantities reflect the amount of unsaturated magnetization even at 2 T.

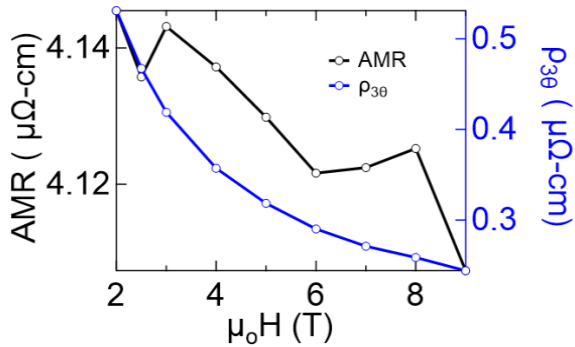

Figure S1 Comparison of the $cos3\theta$ term in the Hall effect and Anisotropic Magnetoresistance (AMR) as a function of magnetic field at 300 K.

**Section 2: Nearly isotropic behavior in the kagome plane**

Fig. S2(b) and (e) show the angular dependence of the Hall effect at 2 K obtained upon field rotation within the *a-c* plane and $a_\perp$-*c* plane, respectively, as illustrated in Fig. S2(a) and (d). Overall the angular behavior of the Hall effect is isotropic within the basal plane as shown in Fig. S2 (b) and (e)[2]. Fig. S2(c) and (f) show the intersection of the curves at different field values close to 90 deg. In Fig. S2(f), various curves intersect at slightly below zero, while in Fig. S2(c) all the curves intersect at zero. The non-zero intersection point in Fig. S2(f) represents the antisymmetric planar Hall effect in $Fe_3Sn_2$.[3]. If there was no antisymmetric planar Hall effect, the curves in Fig. S2(f) at 90° would intersect at zero, i.e., would measure no Hall resistance.

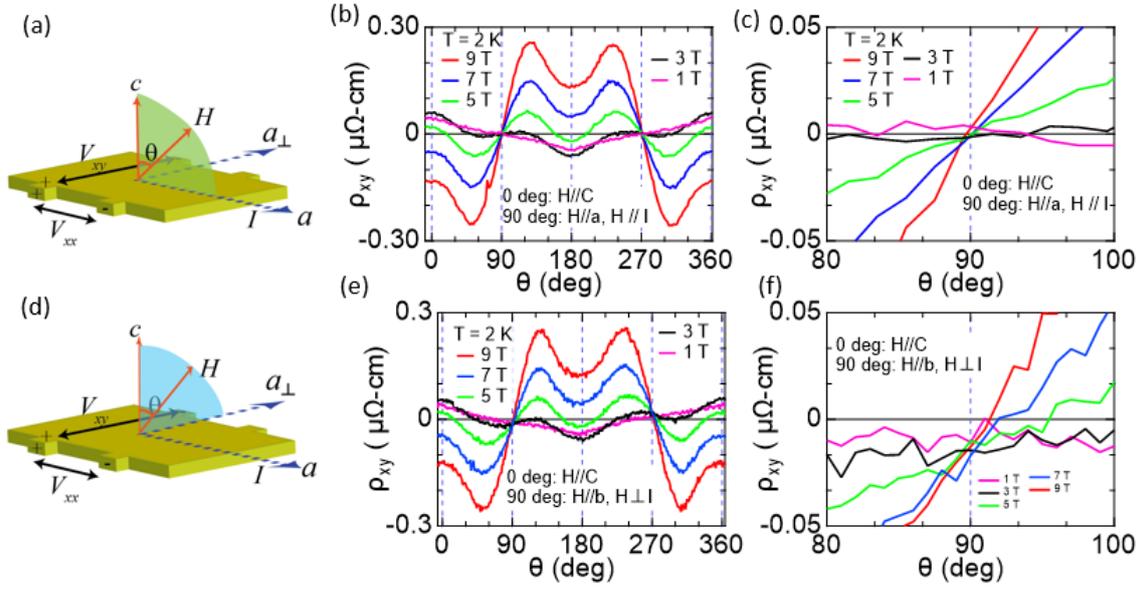

Figure S2 Comparison of the angular dependence of the Hall effect obtained upon field rotation along two different planes at 2 K. (a-c) rotation in a-c plane, (d-f) rotation in $a_\perp$-c plane. (c) and (f) are zoomed versions of (b) and (e) respectively, focusing on the behavior near 90 degrees. For H// $a_\perp$, various field curves cross each other slightly below 0 indicating the presence of a planar Hall effect.

**Section 3: Two band model**

An attempt was made to simultaneously fit the experimental longitudinal resistivity and Hall effect data with the following equations based on a two-carrier model and adding the contribution from the anomalous Hall effect to the Hall channel.

$$\rho_{xx}(H) = \frac{1}{e}\frac{(n_e\mu_e+n_h\mu_h) + (n_h\mu_e+n_e\mu_h)\mu_e\mu_h H^2}{(n_e\mu_e+n_h\mu_h)^2 + [(n_h-n_e)\mu_e\mu_h]^2 H^2}$$

$$\rho_{xy}(H) = \frac{H}{e}\frac{(n_h\mu_h^2-n_e\mu_e^2) + (n_h-n_e)\mu_e^2\mu_h^2 H^2}{(n_e\mu_e+n_h\mu_h)^2 + [(n_h-n_e)\mu_e\mu_h]^2 H^2} + R_A M_Z$$

In the high field limit, $\rho_{xy}(H) \approx \frac{H}{e}\frac{1}{(n_h-n_e)} + R_A M_Z$.

In the low field limit, $\rho_{xy}(H) \approx \frac{H}{e}\frac{(n_h\mu_h^2-n_e\mu_e^2)}{(n_e\mu_e+n_h\mu_h)^2}$.

In the zero field limit, $\rho_{xx}(0) = \frac{1}{e}\frac{1}{(n_e\mu_e+n_h\mu_h)}$.

In the low field limit, $\rho_{xx}(H) \approx \frac{1}{e}\frac{(n_e\mu_e+n_h\mu_h)+(n_h\mu_e+n_e\mu_h)\mu_e\mu_h H^2}{(n_e\mu_e+n_h\mu_h)^2} = \frac{1}{e}\frac{1}{(n_e\mu_e+n_h\mu_h)} + \frac{1}{e}\frac{(n_h\mu_e+n_e\mu_h)\mu_e\mu_h H^2}{(n_e\mu_e+n_h\mu_h)^2}$.

When fitting the experimental data at 2 K to obtain the carrier density and mobility, $\Delta n$ was kept at $1.74 \times 10^{28} m^{-3}$, which is obtained independently, from the slope of the Hall effect at high magnetic fields. The experimental Hall effect as a function of magnetic field fits well with a certain range of mobility and carrier density parameters while the same parameters do not fit well the measured longitudinal resistivity $\rho_{xx}$ as a function of magnetic field. This is demonstrated in the left two curves of Fig. S3. In the first curve, the fitting yields a very small $n_h = 5.5 \times 10^{25} m^{-3}$ but large $n_e = 1.76 \times 10^{28} m^{-3}$, while at the same time a very large $\mu_h = 1.8 \times 10^{-1} m^2 V^{-1} s^{-1}$ but a small $\mu_e = 7.5 \times 10^{-3} m^2 V^{-1} s^{-1}$. $\rho_{xx}(0)$ in this case is $4.3\ \mu\Omega.cm$ which is close to the experimental value of $\rho_{xx}(0) = 4.35\ \mu\Omega.cm$. However, the field dependence of $\rho_{xx}$ is not well fitted. In the second curve, the magnetoresistance (MR) was fit with a power law $H^p$ and using the coefficient of $H^p$ as $(\mu_e \mu_h)^p$, yields $\rho_{xx}(0) = 2.27\ \mu\Omega.cm$, $n_h = 1.69 \times 10^{26} m^{-3}$, $n_e = 1.76 \times 10^{28} m^{-3}$, and $\mu_e = 1.4 \times 10^{-2} m^2 V^{-1} s^{-1}$, $\mu_h = 1.9 \times 10^{-1} m^2 V^{-1} s^{-1}$, where $n_e$ and $\mu_h$ remain almost the same as before. While the Hall effect data fits well, the MR data does not.

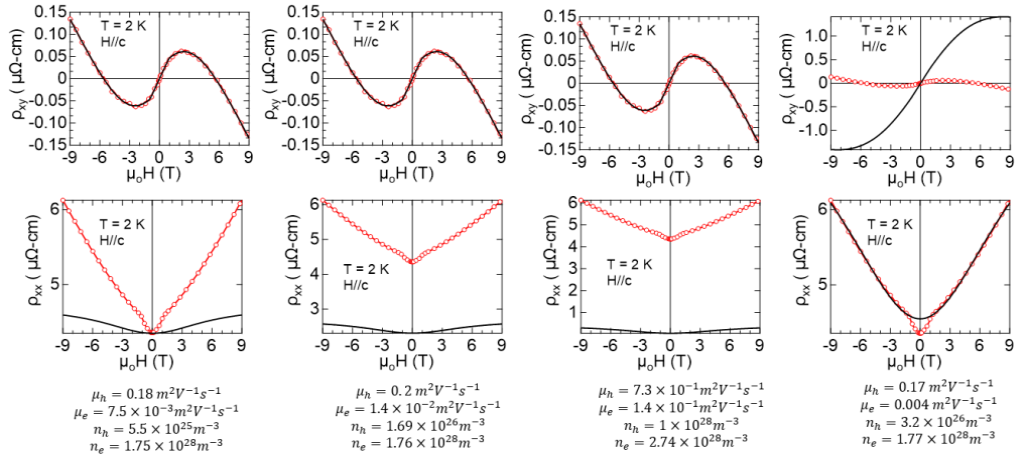

Figure S3 Calculation of Hall and MR effect for various set of parameters.

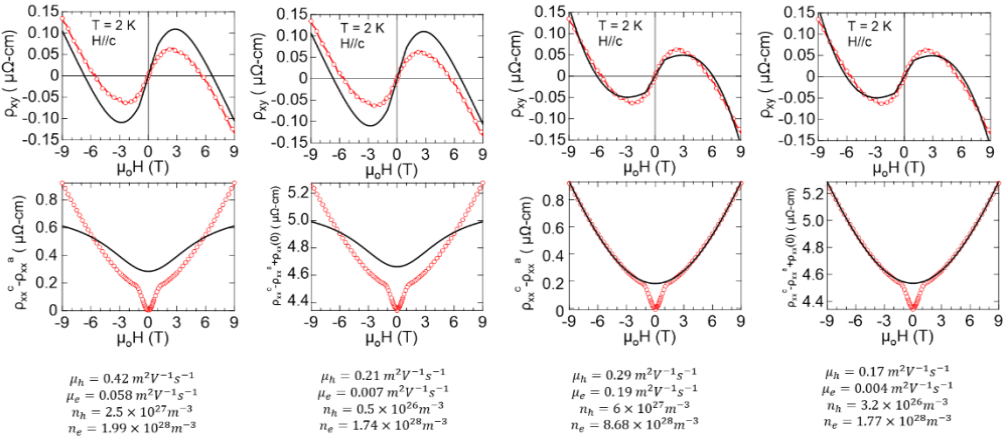

Figure S4 Calculation of adjusted Hall and MR effect for various set of parameters.

Since the large disparity in the hole and electron parameters are unphysical, we tried a different approach, where we put a constraint that the hole and electron carrier density are similar. This is shown in the third curve in Fig. S3. In this scenario, $n_h = 1 \times 10^{28} m^{-3}$, $n_e = 2.74 \times 10^{28} m^{-3}$, $\mu_e = 1.4 \times 10^{-1} m^2 V^{-1} s^{-1}$, and $\mu_h = 7.3 \times 10^{-1} m^2 V^{-1} s^{-1}$ is obtained, which results in a very small value for $\rho_{xx}(0) = 0.055\ \mu\Omega.cm$, two orders of magnitude smaller than the experimental value.

On the other hand, we attempted to fit the longitudinal resistivity by varying the carrier density and mobility. As shown in the fourth curve of Fig. S3, the fitting yields a very small $n_h = 3.2 \times 10^{26} m^{-3}$ but large $n_e = 1.77 \times 10^{28} m^{-3}$, while at the same time a very large $\mu_h = 1.7 \times 10^{-1} m^2 V^{-1} s^{-1}$ but a small $\mu_e = 4 \times 10^{-3} m^2 V^{-1} s^{-1}$. Even though the obtained fitting parameters, which closely fit to the longitudinal resistivity as a function of magnetic field, are not so different from those for Curve 1, they are sufficiently different such that they do not match the experimentally measured Hall effect versus magnetic field.

Furthermore, we attempted to fit by considering only the difference in out of plane and in plane magnetoresistance as the longitudinal resistance to eliminate the spin effect as shown in the first and third curves of Fig. S4 or this difference together with the zero field resistance as in the second and fourth curves of Fig. S4. But in neither of these, a satisfactory fitting could be obtained.

**Section 4: Dependence of carrier density on magnetization direction**

We analyzed the Hall effect in Sample 1 with both a simple single band picture and the two-carrier model as shown in Fig. S5(a) and (b), respectively. In both cases, we see a modification of the carrier density as the magnetization direction changes. This also effectively changes the carrier density vs temperature behavior as shown in Fig. S5(c) and (d) for the single carrier and two carrier model, respectively.

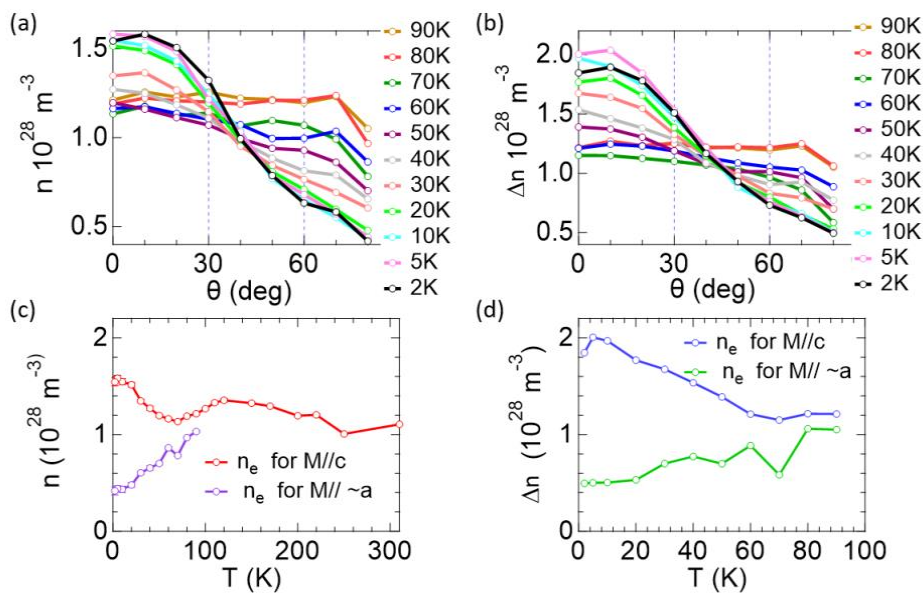

Figure S5 Change in carrier density with the magnetization angle at several temperatures for (a) one carrier model, (b) two carrier model. (c) and (d) Carrier density as a function of temperature for M//c and M close to a.

**Section 5: Additional samples**

To confirm the behavior we observe, we measured two more samples. Both of the additional samples show the same qualitative behavior while the exact value of the Hall effect, mobility, or carrier densities varies. Fig. S6(a) shows the field dependent Hall effect for Sample 2 at 2 K for various field directions in the $ac$ plane, with $\theta = 0$ and $\theta = 90°$ corresponding to $H \parallel c$ and $H \parallel I \parallel a$, respectively. Fig. S6(b) shows the temperature dependence of the carrier density for $M \parallel c$ in Sample 2 using both a single carrier and a two-carrier model. Fig. S6(c) shows the carrier density in Sample 2 being dependent on the angle of the magnetization using a two-carrier model.

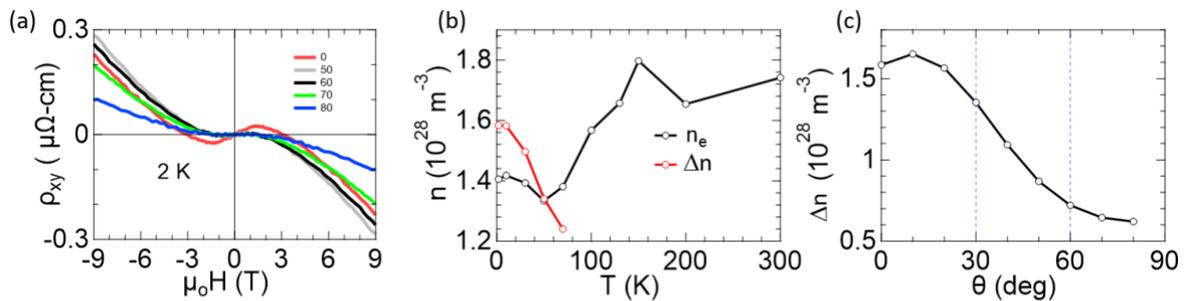

Figure S6 Sample 2: (a) Hall effect at 2 K for various magnetic field directions. Carrier density as a function of (b) temperature, (c) at 2 K, with magnetization angle.

Fig. S7 shows the data for Sample 3. In Fig. S7(a) is plotted the Hall effect with $H \parallel c$ at various temperatures. In Fig. S7(b) the Hall effect at 2 K for the magnetic field at different directions in the *ac* plane is plotted, with $\theta = 0$ and $\theta = 90°$ corresponding to $H \parallel c$ and $H \parallel I \parallel a$, respectively. In Fig. S7(c), the carrier density derived from the Hall effect using a single carrier model is shown for two different magnetization directions with $M \parallel c$ and $M$ almost in the kagome plane. The deviation in carrier density along the different magnetization directions develops at temperatures below 100 K as in Sample 1. In Fig. S7(d), the carrier density depending on the magnetization direction is plotted. It is possible that the sample preparation has an effect on the properties of the sample similar to the case with WTe$_2$[4].

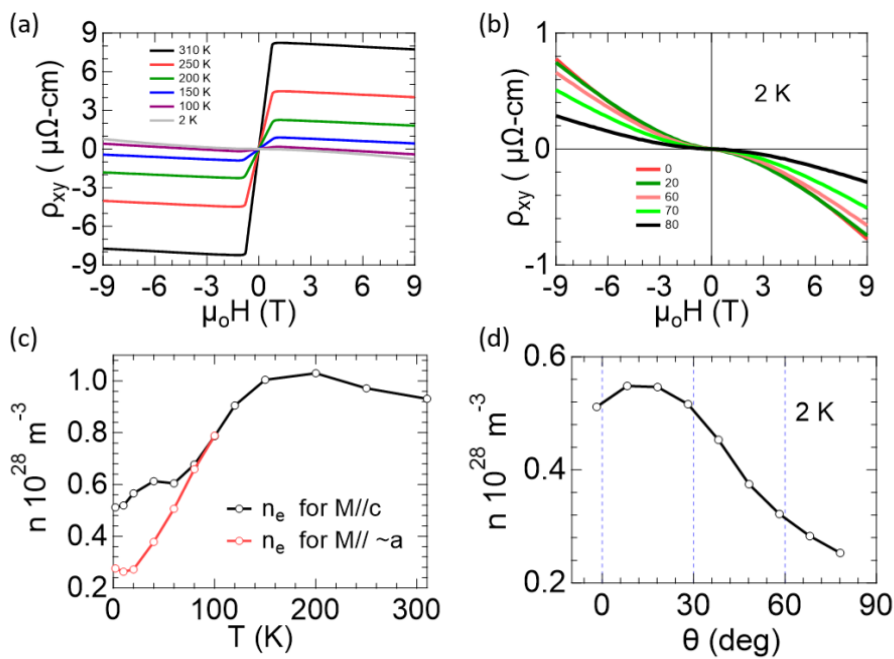

Figure S7 Sample 3: (a) Hall effect at various temperatures with H//c. (b) Hall effect at 2 K with H at various angles. Carrier density as a function of (c) temperature for M//c and M close to a, (d) at 2 K, with magnetization angle.

**Section 6: Comparison of carrier density of various samples**

We compared the evolution of the carrier density with temperature of three different samples. As shown in Fig. S8, in Sample1 and 2, the carrier density increases with decreasing temperature below 100 K, while in Sample 3, it decreases with decreasing temperature.

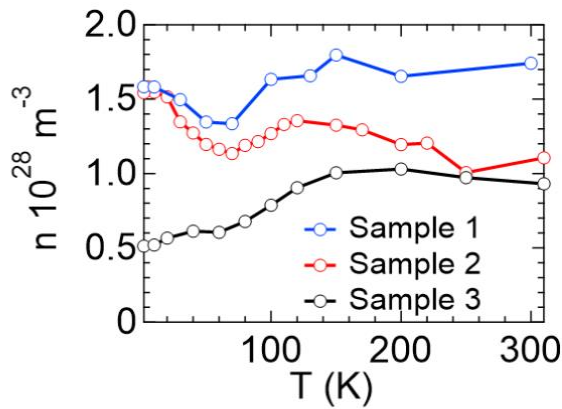

Figure S8 Carrier density vs temperature for the three samples.

## Section 7: Magnetoresistance behavior with magnetic field direction

MR curves gradually change with the magnetization angle at 300 K and 2 K. In these measurements, the field was in the ac plane, with $\theta = 0$ and $\theta = 90°$ corresponding to $H \parallel c$ and $H \parallel I \parallel a$, respectively. At 300 K, the dependence on the magnetization direction is concentrated at the low field due to the dynamics of magnetic domains and associated anisotropic magnetoresistance (AMR)[1], while above saturation, the slope due to the magnon scattering remains nearly constant, independent of the magnetization direction. At 2K, the change in MR depending on the magnetization direction occurs at both low and high field. At low field, we believe the main effect is the dynamics of magnetic domains and associated anisotropic magnetoresistance (AMR)[1]. At high fields, the MR is smaller at field orientations away from the direction normal to the kagome plane. This behavior cannot be solely explained by the reduced orbital magnetoresistance due to the reduced Lorentz force $\propto \vec{j} \times \vec{H} = jH\sin(90° - \theta) = jH\cos\theta$ based on Fig. S9(c), where $\rho_{xx}$ is plotted against $H\cos\theta$. If the change of the magnetoresistance for different field directions were due to the change of the Lorentz force, all the curves would collapse when plotted against $H\cos\theta$, which is not the case. The strong deviation of the MR curves from each other in Fig. S8(c) may be due to the change of the electronic band structure as the magnetization direction is changed.

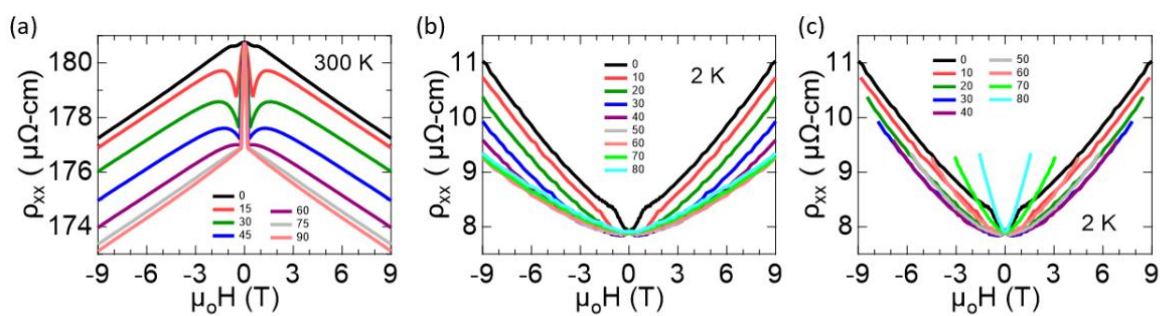

Figure S9 MR of Sample 1 at (a) room temperature, (b) at 2 K for various directions of magnetization. (c) MR at 2 K as a function of only out of plane field component.